\documentclass[aps,pra,preprintnumbers,showpacs,tightenlines]{revtex4}
\usepackage{amssymb}
\usepackage{amsmath}
\usepackage{graphicx}
\usepackage{epsfig}
\usepackage{subfigure}
\usepackage{amsfonts}
\usepackage{CJK}

\begin{document}

\title{Quantum information transfer with superconducting flux qubits coupled to a
resonator}

\author{Chui-Ping Yang}

\address{Department of Physics, Hangzhou Normal University,
Hangzhou, Zhejiang 310036, China}
\date{\today}

\begin{abstract}
We propose a way for implementing quantum information transfer
with two superconducting flux qubits, by coupling them to a
resonator. This proposal does not require adjustment of the level
spacings or uniformity in the device parameters. Moreover, neither
adiabatic passage nor a second-order detuning is needed by this
proposal, thus the operation can be performed much faster when
compared with the previous proposals.
\end{abstract}

\pacs{03.67.Lx, 42.50.Dv, 85.25.Cp} \maketitle
\date{\today}

\textit{Introduction.}---Cavity QED with superconducting qubits including superconducting
charge qubits, phase qubits and flux qubits have
been considered as one of the most promising candidates for
quantum information processing. Superconducting qubits have the
features such as design flexibility, large scale integration, and
compatibility to conventional electronics [1-3]. A cavity or
resonator acts as a ``quantum bus" which can mediate
long-distance, fast interaction between distant superconducting
qubits [4]. 

In recent years, there is much interest in quantum
information transfer (QIT). One of its applications is as follows. When performing
quantum information processing in a practical system, after a step of
processing is completed, we need to transfer the state of the operation
qubit to the memory qubit for storage. Then we need to transfer the state
from the memory qubit to the operation qubit when a further step of
processing is needed. Thus, it is an interesting topic to realize quantum
state transfer between qubits. Experimentally, QIT 
has been demonstrated with superconducting phase qubits and transmon qubits
in cavity QED [5,6]. However, to the best of our knowledge, no experimental
demonstration of QIT with superconducting flux
qubits in cavity QED has been reported.

Several theoretical methods have been proposed for implementing QIT 
with flux qubits (e.g., SQUID qubits) or charge-flux
qubits based on cavity QED technique [7-12]. These methods are useful for
the physical realization of QIT with flux qubits in
cavity QED. However, these methods have some disadvantages. For instances:
(i) the method presented in [8] requires adjustment of the level spacings
of the devices during the operation; (ii) the methods proposed in [7,9-11]
require slowly changing the Rabi frequencies to satisfy the adiabatic
passage; and (iii) the approach introduced in [12] requires a second-order
detuning to achieve an off-resonant Raman coupling between two relevant
levels. Note that the adjustment of the level spacings during the operation
is undesirable and also may cause extra decoherence. In addition, when the
adiabatic passage or a second-order detuning is applied, the operation
becomes slow (the operation time required for the information transfer is on
the order of one microsecond to a few microseconds [7,12]).

In this paper, we present a way for implementing QIT
with two flux qubits coupled to a superconducting resonator. As
shown below, this proposal has the following advantages: (a) the qubits are
not required to have identical level spacings, therefore superconducting
devices, which often have considerable parameter nonuniformity, can be used;
(b) the method does not require adjustment of the level spacings of each
qubit during the operation, thus decoherence caused by tuning the level
spacings is avoided; (c) neither adiabatic passage nor a second-order
detuning is needed, thus the operation is speeded up (as shown below, the
operation time for the information transfer is on the order of ten
nanoseconds).

\textit{Basic theory.}---The flux qubits throughout this paper have three
levels $\left| 0\right\rangle ,$ $\left| 1\right\rangle ,$ and $\left|
2\right\rangle ,$ which form a $\Lambda $-type configuration depicted in
Fig. 1. The transition between the two lowest levels is forbidden due to the
optical selection rules [13] or weak via increasing the potential barrier
between the two levels $\left| 0\right\rangle $ and $\left| 1\right\rangle $
[14-16]. The qubits with this three-level structure could be a
radio-frequency superconducting quantum interference device (rf SQUID)
consisting of one Josephson junction enclosed by a superconducting loop [see
Fig.~2(a)], or a superconducting device with three Josephson junctions
enclosed by a superconducting loop [Fig. 2(b)]. When the loop is flux-biased
properly and/or the device parameters are chosen appropriately, the desired
level structure shown in Fig. 1 is available. For flux qubits, the
two logic states of a qubit are represented by the two lowest levels $\left|
0\right\rangle $ and $\left| 1\right\rangle .$

\begin{figure}[tbp]
\begin{center}
\includegraphics[bb=88 368 260 536, width=6.6 cm, clip]{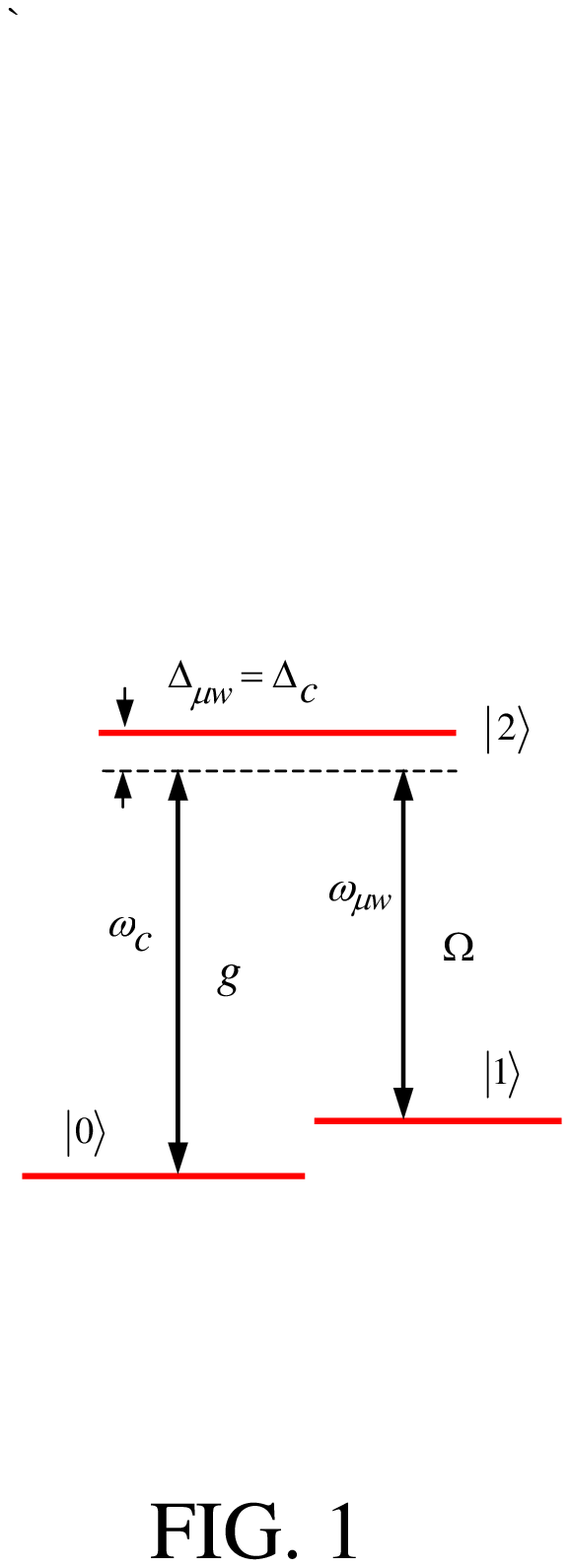} %
\vspace*{-0.08in}
\par
\end{center}
\caption{(Color online) Qubit-resonator-pulse resonant Raman coupling.
The tunneling between the two lowest levels is forbidden or weak,
such that quantum information for each qubit, encoded in the two lowest
levels, can be stored for a long time.}
\label{fig:1}
\end{figure}

\textit{A). Qubit-resonator-pulse resonant Raman coupling.} Consider a flux
qubit coupled to a single-mode resonator and driven by a classical microwave
pulse (Fig.~1). Suppose that the resonator mode is coupled to the $\left|
0\right\rangle \leftrightarrow \left| 2\right\rangle $ transition but
decoupled (highly detuned) from the transition between any other two levels.
In addition, assume that the classical microwave pulse is coupled to the $%
\left| 1\right\rangle \leftrightarrow \left| 2\right\rangle $ transition but
decoupled from the transition between any other two levels. The Hamiltonian
of the system can thus be written as
\begin{equation}
H=\sum_lE_l\sigma _{ll}+\omega _ca^{+}a+\hbar g(a^{+}\sigma _{02}^{-}+\text{%
H.c.})+\hbar \Omega (e^{i\omega _{\mu w}t}\sigma _{12}^{-}+\text{H.c.}),
\end{equation}
where $a^{+}$ and $a$ are the photon creation and annihilation operators of
the resonator mode with frequency $\omega _c$; $g$ is the coupling constant
between the resonator mode and the $\left| 0\right\rangle \leftrightarrow
\left| 2\right\rangle $ transition of the qubit; $\Omega $ is the Rabi
frequency of the pulse and $\omega _{\mu w}$ is the frequency of the pulse; $%
\sigma _{02}^{-}=\left| 0\right\rangle \left\langle 2\right| ,$ $\sigma
_{12}^{-}=\left| 1\right\rangle \left\langle 2\right| ,$ and $\sigma
_{ll}=\left| l\right\rangle \left\langle l\right| $ ($l=0,1,2$).

Suppose that the resonator mode is off-resonant with the $\left|
0\right\rangle \leftrightarrow \left| 2\right\rangle $ transition, i.e., $%
\Delta _c=\omega _{02}-\omega _c\gg g,$ and the pulse is off-resonant with
the $\left| 1\right\rangle \leftrightarrow \left| 2\right\rangle $
transition, i.e., $\Delta _{\mu w}=\omega _{12}-\omega _{\mu w}\gg \Omega $
(Fig.~1), where $\omega _{02}$ is the $\left| 0\right\rangle \leftrightarrow
\left| 2\right\rangle $ transition frequency and $\omega _{12}$ is the $%
\left| 1\right\rangle \leftrightarrow \left| 2\right\rangle $ transition
frequency. Under this condition, the level $\left| 2\right\rangle $ can be
adiabatically eliminated [17]. Thus, for $\Delta _c=\Delta _{\mu w},$ the
effective Hamiltonian in the interaction picture is [12,18]
\begin{equation}
H_{\text{eff}}=-\hbar \left[ \frac{\Omega ^2}{\Delta _{\mu w}}\sigma _{11}+%
\frac{g^2}{\Delta _c}a^{+}a\sigma _{00}+\frac{\Omega g}{\Delta _c}%
(a^{+}\sigma _{01}^{-}+\text{H.c.})\right] ,
\end{equation}
where $\sigma _{01}^{-}=\left| 0\right\rangle \left\langle 1\right| $. The
first two terms in Eq.~(2) are ac-Stark shifts of the levels $\left|
1\right\rangle $ and $\left| 0\right\rangle $ induced by the pulse and the
resonator mode, respectively; while the last two terms in Eq.~(2) are the
familiar Jaynes-Cummings interaction, describing the resonant Raman coupling
between the two lowest levels $\left| 0\right\rangle $ and $\left|
1\right\rangle ,$ which results from the cooperation of the resonator mode
and the pulse.

For the case of $\Omega =g,$ the initial states $\left| 0\right\rangle
\left| 1\right\rangle _c$ and $\left| 1\right\rangle \left| 0\right\rangle
_c $ of the system, under the Hamiltonian (2), evolve as follows
\begin{eqnarray}
\left| 0\right\rangle \left| 1\right\rangle _c &\rightarrow &e^{ig^2t/\Delta
_c}\left[ \cos \left( g^2t/\Delta _c\right) \left| 0\right\rangle \left|
1\right\rangle _c-i\sin \left( g^2t/\Delta _c\right) \left| 1\right\rangle
\left| 0\right\rangle _c\right] ,  \nonumber \\
\left| 1\right\rangle \left| 0\right\rangle _c &\rightarrow &e^{ig^2t/\Delta
_c}\left[ -i\sin \left( g^2t/\Delta _c\right) \left| 0\right\rangle \left|
1\right\rangle _c+\cos \left( g^2t/\Delta _c\right) \left| 1\right\rangle
\left| 0\right\rangle _c\right] ,
\end{eqnarray}
where $\left| 0\right\rangle _c$ and $\left| 1\right\rangle _c$ are the
vacuum state and the single-photon state of the resonator mode,
respectively. The state $\left| 0\right\rangle \left| 0\right\rangle _c$
remains unchanged under the Hamiltonian~(2).

The coupling strength $g$ may vary with different qubits due to non-uniform
device parameters and/or non-exact placement of qubits in the resonator.
Therefore, in the information transfer operation below, we
will replace $g$ by $g_a$ and $g_b$ for qubits $a$ and $b,$ respectively.
Accordingly, we will replace $\Delta _c$ by $\Delta _c^a$ and $\Delta _c^b,$
$\Delta _{\mu w}$ by $\Delta _{\mu w}^a$ and $\Delta _{\mu w}^b,$ and $%
\Omega $ by $\Omega _a$ and $\Omega _b$ for qubits $a$ and $b,$ respectively.

\textit{B). Qubit-pulse resonant interaction.} Consider a three-level flux
qubit driven by a classical microwave pulse. Suppose that the pulse is
resonant with the transition $\left| i\right\rangle \leftrightarrow \left|
j\right\rangle $ of the qubit but decoupled from the transition between any
two other levels. Here, $\left| i\right\rangle $ is the lower energy level.
The interaction Hamiltonian in the interaction picture is given by
\begin{equation}
H_I=\hbar \left( \widetilde{\Omega }e^{i\phi }\left| i\right\rangle
\left\langle j\right| +\text{H.c.}\right) ,
\end{equation}
where $\widetilde{\Omega }$ and $\phi $ are the Rabi frequency and the
initial phase of the pulse, respectively. Based on the Hamiltonian (4), it
is straightforward to show that a pulse of duration $t$ results in the
following rotation
\begin{eqnarray}
\left| i\right\rangle  &\rightarrow &\cos \widetilde{\Omega }t\left|
i\right\rangle -ie^{-i\phi }\sin \widetilde{\Omega }t\left| j\right\rangle ,
\nonumber \\
\left| j\right\rangle  &\rightarrow &\cos \widetilde{\Omega }t\left|
j\right\rangle -ie^{i\phi }\sin \widetilde{\Omega }t\left| i\right\rangle .
\end{eqnarray}

The transition frquency $\omega _{ij}$ between the two levels $\left|
i\right\rangle $ and $\left| j\right\rangle $ may be different for qubits $a$
and $b$ due to their nonidentical level spacings. Thus, in the following, we
will replace $\omega _{ij}$ by $\omega _{ij}^a$ and $\omega _{ij}^b$ for
qubits $a$ and $b$, respectively.

\textit{Quantum information transfer.}---Let us now consider two
superconducting flux qubits $a$ and $b$ coupled to a resonator [Fig.~2(c)].
Each qubit has a $\Lambda $-type three-level configuration as depicted in
Fig. 1. The quantum information of a qubit is encoded by the two lowest
levels $\left| 0\right\rangle $ and $\left| 1\right\rangle $. Suppose that
qubit $a$ is the original carrier of quantum information, which is in an
arbitrary state $\alpha \left| 0\right\rangle +\beta \left| 1\right\rangle .$
The QIT from qubit $a$ to qubit $b$ initially in the
state $\left| 1\right\rangle $ is described by
\begin{equation}
\left( \alpha \left| 0\right\rangle _a+\beta \left| 1\right\rangle _a\right)
\left| 1\right\rangle _b\rightarrow \left| 1\right\rangle _a\left( \alpha
\left| 0\right\rangle _b+\beta \left| 1\right\rangle _b\right) .
\end{equation}
From Eq. (6), one can see that this process can be done via a transformation
that satisfies the following truth table:
\begin{eqnarray}
\left| 0\right\rangle _a\left| 1\right\rangle _b &\rightarrow &\left|
1\right\rangle _a\left| 0\right\rangle _b,  \nonumber \\
\left| 1\right\rangle _a\left| 1\right\rangle _b &\rightarrow &\left|
1\right\rangle _a\left| 1\right\rangle _b.
\end{eqnarray}

\begin{figure}[tbp]
\begin{center}
\includegraphics[bb=29 301 534 701, width=9.6 cm, clip]{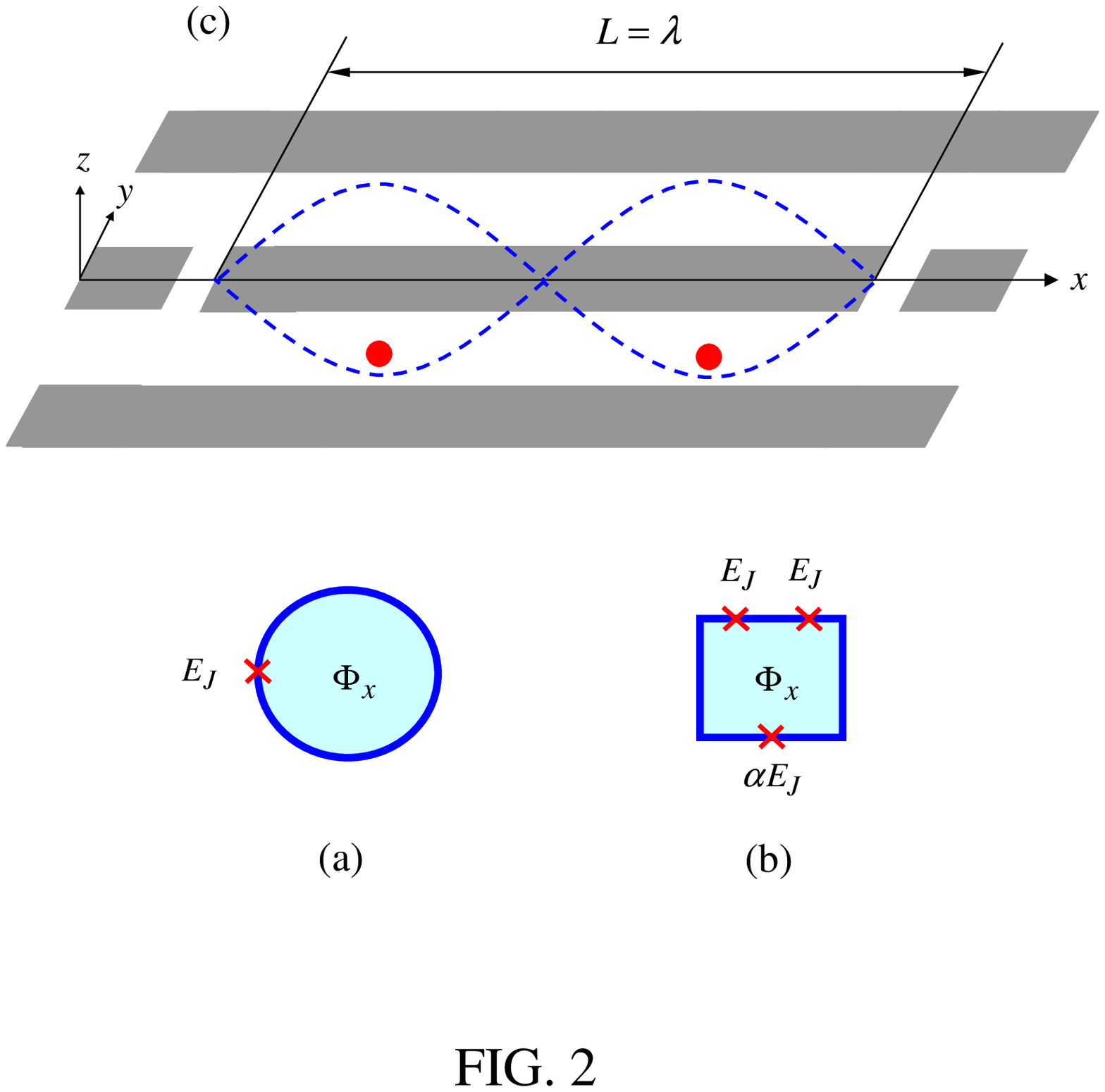} %
\vspace*{-0.08in}
\end{center}
\caption{(Color online) (a) An rf SQUID consisting of one Josephson junction
enclosed by a superconducting loop. (b) A superconducting device with three
Josephson junctions enclosed by a loop. The level spacings of a flux qubit
shown in Fig.~1 can be adjusted by changing external magnetic flux $\Phi_x$
applied to the loop. Here, $E_{J}$ is the Josephson junction energy, and $%
0<\alpha<1$. (c) Sketch of the setup for two superconducting flux qubits
(red circles) and a (grey) standing-wave quasi-one-dimensional coplanar
waveguide resonator. The two blue curved lines represent the standing wave
magnetic field, which is in the $z$ direction. Each qubit could be an rf
SQUID shown in (a) or a superconducting device with three Josephson
junctions shown in (b). The qubits are placed at antinodes of the resonator
mode to achieve maximal qubit-resonator coupling constants. The
superconducting loop of each qubit is located in the plane of the resonator
between the two lateral ground planes (i.e., the $x$-$y$ plane). $\lambda$
is the wavelength of the resonator mode and $L$ is the length of the
resonator.}
\label{fig:2}
\end{figure}

To realize the transformation (7), suppose that the resonator mode is
off-resonant with the $\left| 0\right\rangle \leftrightarrow \left|
2\right\rangle $ transition of each qubit (with a detuning $\Delta
_c^a=\omega _{02}^a-\omega _c$ for qubit $a$ while $\Delta _c^b=\omega
_{02}^b-\omega _c$ for qubit $b$) but highly detuned (decoupled) from the
transition between any other two levels of each qubit. Note that this
condition can be readily achieved by prior adjustment of the level spacings
of the qubits before the operation (e.g., this is doable for superconducting
qubits by varying the external flux applied to the superconducting loop
[14-16,19,20]). In addition, we assume that the resonator mode is initially
in the vacuum state $\left| 0\right\rangle _c.$

We find that the transformation (7) can be implemented through the following
operations:

\begin{figure}[tbp]
\begin{center}
\includegraphics[bb=105 122 476 724, width=10.6 cm, clip]{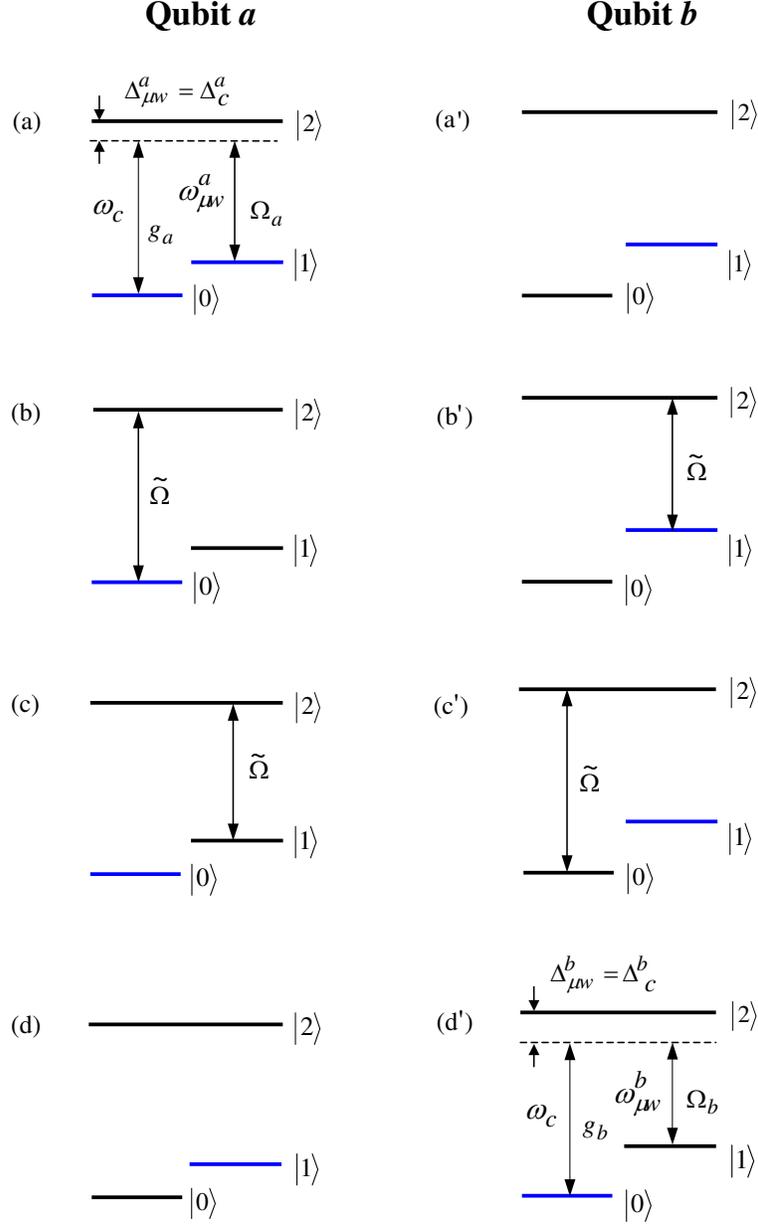} %
\vspace*{-0.08in}
\end{center}
\caption{(Color online) Illustration of qubits interacting with the
resonator mode and/or the microwave pulses for each step of operations
during the information transfer operation. Figures from top to bottom
correspond to the operations of steps (i)$\sim $(iv), respectively. The
figures on the left side correspond to qubit $a$ while the figures on the
right side correspond to qubit $b.$ In addition, in each figure, the blue
lines represent the level population of qubits before each step of
operation. Note that the nonidentical level spacings for the two qubits are
caused by the nonuniform device parameters of the two qubits.}
\label{fig:3}
\end{figure}

Step (i): Apply a microwave pulse (with a frequency $\omega _{\mu w}^a$) to
qubit $a$ [Fig. 3(a)]. The pulse is off-resonant with the $\left|
1\right\rangle \leftrightarrow \left| 2\right\rangle $ transition of qubit $%
a,$ with a detuning $\Delta _{\mu w}^a=\omega _{12}^a-\omega _{\mu w}^a$. To
establish the resonant Raman coupling between the two levels $\left|
0\right\rangle $ and $\left| 1\right\rangle ,$ set $\Delta _{\mu w}^a=$ $%
\Delta _c^a.$ The Rabi frequency $\Omega _a$ of the pulse is set by $\Omega
_a=g_a$, which can be achieved by adjusting the pulse intensity. It can be
seen from Eq. (3) that after a pulse duration $t_1=\pi \Delta _c^a/\left(
2g_a^2\right) ,$ the state $\left| 1\right\rangle _a\left| 0\right\rangle _c$
for qubit $a$ and the resonator mode is transformed to the state $\left|
0\right\rangle _a\left| 1\right\rangle _c.$ On the other hand, the state $%
\left| 0\right\rangle _a\left| 0\right\rangle _c$ remains unchanged during
the pulse.

To have a qubit coupled with the resonator mode, the qubit must be in either
of the levels $\left| 0\right\rangle $ or the level $\left| 2\right\rangle .$
Note that qubit $b$ was initially prepared in the state $\left|
1\right\rangle $ and kept in the same state $\left| 1\right\rangle $ during
this step. Therefore, qubit $b$ is decoupled from the resonator mode during
this step.

Step (ii): Apply a microwave pulse (with a frequency $\omega _{\mu
w}^a=\omega _{02}^a$ and a phase $\phi =-\pi /2$) to qubit $a$ [Fig. 3(b)]
and a microwave pulse (with a frequency $\omega _{\mu w}^b=\omega _{12}^b$
and a phase $\phi =-\pi /2$) to qubit $b$ [Fig. 3(b$^{\prime }$)]. The Rabi
frequency for each pulse is $\widetilde{\Omega }$. Thus, it can be seen from
Eq.~(5) that after the pulse duration $t_2=\pi /(2\widetilde{\Omega }),$ the
state $\left| 0\right\rangle $ of qubit $a$ is transformed to the state $%
\left| 2\right\rangle $ while the state $\left| 1\right\rangle $ of qubit $b$
is transformed to the state $\left| 2\right\rangle $.

Step (iii): Apply a microwave pulse (with a frequency $\omega _{\mu
w}^a=\omega _{12}^a$ and a phase $\phi =\pi /2$) to qubit $a$ [Fig. 3(c)]
and a microwave pulse (with a frequency $\omega _{\mu w}^b=\omega _{02}^b$
and a phase $\phi =\pi /2$) to qubit $b$ [Fig. 3(c$^{\prime }$)]. The Rabi
frequency for each pulse is $\widetilde{\Omega }$. It can be found from
Eq.~(5) that after the pulse duration $t_3=\pi /(2\widetilde{\Omega }),$ the
state $\left| 2\right\rangle $ of qubit $a$ is transformed to the state $%
\left| 1\right\rangle $ while the state $\left| 2\right\rangle $ of qubit $b$
is transformed to the state $\left| 0\right\rangle $.

Step (iv): Apply a microwave pulse (with a frequency $\omega _{\mu w}^b$) to
qubit $b$ [Fig.~3(d$^{\prime }$)]. The pulse is off-resonant with the $%
\left| 1\right\rangle \leftrightarrow \left| 2\right\rangle $ transition of
qubit $b,$ with a detuning $\Delta _{\mu w}^b=\omega _{12}^b-\omega _{\mu
w}^b=\Delta _c^b$ [Fig. 3(d$^{\prime }$)]. The Rabi frequency $\Omega _b$ of
the pulse is set by $\Omega _b=g_b$. It can be seen from Eq. (3) that after
a pulse duration $t_4=\pi \Delta _c^b/\left( 2g_b^2\right) ,$ the state $%
\left| 0\right\rangle _b\left| 1\right\rangle _c$ for qubit $b$ and the
resonator mode is transformed to the state $\left| 1\right\rangle _b\left|
0\right\rangle _c.$ On the other hand, the state $\left| 0\right\rangle
_b\left| 0\right\rangle _c$ remains unchanged during the pulse.

The states of the whole system after each step of the above operations are
summarized in the following table:

\begin{equation}
\begin{array}{c}
\left| 01\right\rangle \left| 0\right\rangle _c \\
\left| 11\right\rangle \left| 0\right\rangle _c
\end{array}
\stackrel{Step(i)}{\longrightarrow }
\begin{array}{c}
\left| 01\right\rangle \left| 0\right\rangle _c \\
\left| 01\right\rangle \left| 1\right\rangle _c
\end{array}
\stackrel{Step(ii)}{\longrightarrow }
\begin{array}{c}
\left| 22\right\rangle \left| 0\right\rangle _c \\
\left| 22\right\rangle \left| 1\right\rangle _c
\end{array}
\stackrel{Step(iii)}{\longrightarrow }
\begin{array}{c}
\left| 10\right\rangle \left| 0\right\rangle _c \\
\left| 10\right\rangle \left| 1\right\rangle _c
\end{array}
\stackrel{Step(iv)}{\longrightarrow }
\begin{array}{c}
\left| 10\right\rangle \left| 0\right\rangle _c \\
\left| 11\right\rangle \left| 0\right\rangle _c
\end{array}
\end{equation}
where $\left| ij\right\rangle $ is abbreviation of the state $\left|
i\right\rangle _a\left| j\right\rangle _b$ of qubits ($a,b$) with $i,j\in
\{0,1,2\}$. It can be found from Eq.~(8) that the transformation (7) was
achieved with two qubits after the above process. Namely, the information
originally carried by qubit $a$ is transferred to qubit $b$ while the
resonator mode returns to its original vacuum state.

From the description above, it can be seen that:

(a) Compared with the previous proposal [8], the method presented here does
not require adjustment of the level spacings of each qubit during the
operation, and thus decoherence caused by the tuning of the level spacings
of the qubits is avoided;

(b) Compared with the previous proposals [7,9-11], the present method does
not require slow variation of the Rabi frequency, and thus the operation is
speeded up;

(c) In contrast to the previous proposal [12], this method does not require
a finite second-order detuning $\delta _a=\Delta _c^a-\Delta _{\mu w}^a$ or $%
\delta _b=\Delta _c^b-\Delta _{\mu w}^b,$ and thus the operation can be
performed faster by one order;

(d) The level spacings for the two qubits do not need to be identical,
therefore the nonuniformity in the device parameters is tolerable.

\textit{Discussion.}---The occupation probability $p_{2,a}$ of the level $%
\left| 2\right\rangle $ for qubit $a$ during step (i) and the occupation
probability $p_{2,b}$ of the level $\left| 2\right\rangle $ for qubit $b$
during step (iv) are given by [20]
\begin{eqnarray}
p_{2,a} &\simeq &\frac 12\left( \frac{4\Omega _a^2}{4\Omega _a^2+\left(
\Delta _{\mu w}^a\right) ^2}+\frac{4g_a^2}{4g_a^2+\left( \Delta _c^a\right)
^2}\right) ,  \nonumber \\
p_{2,b} &\simeq &\frac 12\left( \frac{4\Omega _b^2}{4\Omega _b^2+\left(
\Delta _{\mu w}^b\right) ^2}+\frac{4g_b^2}{4g_b^2+(\Delta _c^b)^2}\right) .
\end{eqnarray}
The occupation probabilities $p_{2,a}$ and $p_{2,b}$ need to be negligibly
small in order to reduce the operation error. For the choice of $\Delta
_{\mu w}^a=\Delta _c^a=10g_a,\Delta _{\mu w}^b=\Delta _c^b=10g_b,$ $\Omega
_a=g_a,$ and $\Omega _b=g_b$, we have $p_{2,a},$ $p_{2,b}\sim 0.04$, which
can be further reduced by increasing the ratio of $\Delta _c^a/g_a,$ $\Delta
_c^a/g_b,$ $\Omega _a/\Delta _a,$ and $\Omega _b/\Delta _b.$

The level $\left| 2\right\rangle $ of each qubit is only occupied in steps
(ii) and (iii). Because resonant pulses are applied in these steps, the
pulse durations $t_2$ for step (ii) and $t_3$ for step (iii) can be reduced
by increasing the pulse Rabi frequencies, such that $t_2\ll T_2$ and $t_3\ll
T_2$ (where $T_2$ is the spontaneous time of the level $\left|
2\right\rangle $ of the qubits). In this way, spontaneous emission from the
level $\left| 2\right\rangle $ can be suppressed.

The levels $\left| 0\right\rangle $ and $\left| 2\right\rangle $ of qubit $a$
are populated during the pulse for step (ii) [Fig. 3(b)] and the level $%
\left| 2\right\rangle $ of qubit $a$ is populated during the pulse for step
(iii) [Fig. 3(c)]. Thus, when the resonator mode is in the single-photon
state $\left| 1\right\rangle _c,$ the off-resonant interaction between the
resonator mode and the $\left| 0\right\rangle \leftrightarrow \left|
2\right\rangle $ transition of  qubit $a$ induces a phase shift $\exp \left(
it_2g_a^2/\Delta _c^a\right) $ $\left[ \exp \left( -it_2g_a^2/\Delta
_c^a\right) \right] $ to the state $\left| 0\right\rangle $ $\left( \left|
2\right\rangle \right) $ of qubit $a$ for step (ii) and $\exp
(-it_3g_a^2/\Delta _c^a)$ to the state $\left| 2\right\rangle $ of qubit $a$
for step (iii)$.$ In addition, the level $\left| 2\right\rangle $ of qubit $b
$ is populated during the pulse for step (ii) [Fig. 3(b$^{\prime }$)] and
the levels $\left| 0\right\rangle $ and $\left| 2\right\rangle $ of qubit $b$
are populated during the pulse for step (iii) [Fig. 3(c$^{\prime }$)].
Hence, when the resonator mode is in the single-photon state $\left|
1\right\rangle _c,$ the off-resonant interaction between the resonator mode
and the $\left| 0\right\rangle \leftrightarrow \left| 2\right\rangle $
transition of qubit $b$ induces a phase shift $\exp \left( -it_2g_b^2/\Delta
_c^b\right) $ to the state $\left| 2\right\rangle $ of qubit $b$ for step
(ii) and $\exp (it_3g_b^2/\Delta _c^b)$ $[\exp (-it_3g_b^2/\Delta _c^b)]$ to
the state $\left| 0\right\rangle $ $\left( \left| 2\right\rangle \right) $
of qubit $b$ for step (iii)$.$ These phase shifts, which are not considered
in Eq. (8), will affect the desired information transfer performance.
However, note that $t_2,$ $t_3\propto 1/\widetilde{\Omega }.$ Thus, these
unwanted phase shifts can be made negligibly small, by increasing the pulse
Rabi frequencies such that $\widetilde{\Omega }\gg g_a^2/\Delta _c^a,$ $%
g_b^2/\Delta _c^b$. To see this more clearly, we will give an analysis on
the effect of the unwanted qubit-resonator off-resonant interaction on the
fidelity of the QIT.

In the ideal case, it can be seen from Eq.~(8) that after the operations
described above, the state of the two qubits and the resonator mode is $%
\left| \psi _{id}\left( \tau \right) \right\rangle =\left| 1\right\rangle
_a\left( \alpha \left| 0\right\rangle _b+\beta \left| 1\right\rangle
_b\right) \otimes \left| 0\right\rangle _c.$ Here, $\tau $ is the total
operation time. On the other hand, when the off-resonant interaction between
the resonator mode and the $\left| 0\right\rangle \leftrightarrow \left|
2\right\rangle $ transition of each qubit is included during steps (ii) and
(iii), one can easily work out the expression for the final state $\left|
\psi \left( \tau \right) \right\rangle $ of the whole system after
performing the operations above. To simplify our presentation, we will not
give a complete expression for $\left| \psi \left( \tau \right)
\right\rangle $ due to its complexity. A simple calculation shows that the
fidelity for the QIT is
\begin{eqnarray}
F &=&\left| \left\langle \psi _{id}\left( \tau \right) \right| \left. \psi
\left( \tau \right) \right\rangle \right| ^2  \nonumber \\
&=&\left( \left| \alpha \right| ^2+pq\left| \beta \right| ^2\right) ,
\end{eqnarray}
where
\begin{eqnarray}
p &=&\frac{\widetilde{\Omega }}{\sqrt{\widetilde{\Omega }^2+s_a^2/4}}\sin
\left[ \frac{\pi \sqrt{\widetilde{\Omega }^2+s_a^2/4}}{2\widetilde{\Omega }}%
\right] ,\;  \nonumber \\
q &=&\frac{\widetilde{\Omega }}{\sqrt{\widetilde{\Omega }^2+s_b^2/4}}\sin
\left[ \frac{\pi \sqrt{\widetilde{\Omega }^2+s_b^2/4}}{2\widetilde{\Omega }}%
\right] ,
\end{eqnarray}
with $s_a=2g_a^2/\Delta _c^a$ and $s_b=2g_b^2/\Delta _c^b.$

\begin{figure}[tbp]
\includegraphics[bb=32 239 485 537, width=8.6 cm, clip]{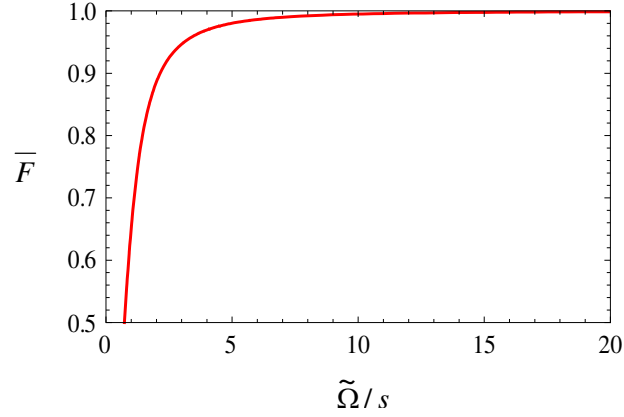} %
\vspace*{-0.08in}
\caption{Average fidelity $\overline{F}$ as a function of the Rabi frequency
$\widetilde{\Omega }$ (in unit of $s$).}
\label{fig:4}
\end{figure}

Defining $\alpha =\cos \frac \vartheta 2$ and $\beta =e^{i\varphi }\sin
\frac \vartheta 2,$ where $\vartheta \in \left[ 0,\pi \right] $ and $\varphi
\in \left[ 0,2\pi \right] .$ Thus, the average fidelity over all possible
initial states of the message qubit $a$ is given by
\begin{eqnarray}
\overline{F} &=&\int_0^{2\pi }d\varphi \int_0^\pi F\sin \vartheta d\vartheta
/4\pi   \nonumber \\
&=&\frac 13\left( 1+p^2q^2+p^4q^4\right) .
\end{eqnarray}
It can be verified that when the unwanted qubit-resonator off-resonant
interaction in steps (ii) and (iii) is not considered (i.e., the case for $%
g_a^2/\Delta _c^a,g_b^2/\Delta _c^b=0$ or $s_a,s_b=0$), we have $p=q=1,$
leading to $F=1$ and $\overline{F}=1.$ We have plotted the average fidelity $%
\overline{F}$ as a function of $\widetilde{\Omega }/s$ for the choice of $%
s_a=s_b=s$ (Fig. 4). One can see from Fig. 4 that the average fidelity $%
\overline{F}$ increases as the pulse Rabi frequency $\widetilde{\Omega }$ becomes
larger, and the $\overline{F}$ is $\sim 1$ when $\widetilde{\Omega }=10s.$
This result demonstrates that the effect of the qubit-resonator off-resonant
interaction in steps (ii) and (iii) on the fidelity of the operation is
negligible when the pulse Rabi frequencies are sufficiently large.

Finally, let us give a rough estimate on the operation time. As shown above,
the total operation time for the information transfer is
\begin{equation}
\tau =\pi \Delta _c^a/\left( 2g_a^2\right) +\pi \Delta _c^b/\left(
2g_b^2\right) +\pi /\widetilde{\Omega }.
\end{equation}
Without loss of generality, let us consider $g_a\sim g_b$ $\sim 3.0\times
10^9$ s$^{-1}$, which is available at present [21]. By choosing $\Delta
_c^a=10g_a,\Delta _c^b=10g_b,$ and $\widetilde{\Omega }\sim 10g_a,$ we have $%
\tau \sim 11$ ns.

\textit{Conclusion.}---We have proposed a way for realizing the quantum
information transfer with superconducting flux qubits coupled to a
resonator. As shown above, this proposal avoids most of the problems
existing in the previous proposals. Finally, it is noted that the method
presented here is quite general, which can be applied to the other physical
systems such as atoms and quantum dots with the $\Lambda $-type three-level
structure within cavity QED.

\textit{Acknowledgments.}---This work was supported by the National Natural
Science Foundation of China under Grant No. 11074062 and the starting
research funds from Hangzhou Normal University.

\end{document}